\documentstyle[12pt]{article}
\newcommand{\cl}{\centerline}
\renewcommand{\theequation}{\arabic{equation}}
\newcommand{\beq}{\begin{equation}}
\newcommand{\eeq}{\end{equation}}
\newcommand{\be}{\begin{equation}}
\newcommand{\ee}{\end{equation}}
\newcommand{\bea}{\begin{eqnarray}}
\newcommand{\eea}{\end{eqnarray}}
\newcommand{\nn}{\nonumber}
\newcommand\pa{\partial}
\newcommand\na{\vec{\nabla}}
\newcommand\un{\underline}

\newcommand\pr{\prime}
\begin{document}
\begin{titlepage}
\setlength{\textwidth}{5.0in} \setlength{\textheight}{7.5in}
\setlength{\parskip}{0.0in} \setlength{\baselineskip}{18.2pt}

\hfill
%{\tt  HD-THEP-02-30, SOGANG-HEP 296/02}
\begin{center}
{\large{{\bf Constraint structure of O(3) nonlinear sigma model
revisited}}}
\end{center}
\vskip 0.5cm

\begin{center}
{Soon-Tae Hong$^{1}$, Yong-Wan Kim$^{2}$, Young-Jai Park$^{2}$ and 
Klaus D. Rothe$^{3}$}
\end{center}
\vskip 0.5cm

\begin{center}
{$^{1}$Department of Science Education\\
Ewha Womans University, Seoul 120-750, Korea}\\
\vskip 0.3cm
{$^{2}$Department of Physics and Basic Science Research Institute,\\
Sogang University, C.P.O. Box 1142, Seoul 100-611, Korea}\\
\vskip 0.3cm
{$^{3}$Institut f\"ur Theoretische Physik,\\
Universit\"at Heidelberg, Philosophenweg 16, D-69120 Heidelberg, Germany}
\end{center}
\vskip 0.5cm

%\date{\today}
%\maketitle

\cl{October 9, 2002}
%\cl{\today}
%\vskip 0.5cm
\vfill
\begin{center}
{\bf ABSTRACT}
\end{center}
\begin{quotation}

We study the constraint structure of the O(3) nonlinear sigma model in the framework 
of the Lagrangian, symplectic, Hamilton-Jacobi as well as the
Batalin-Fradkin-Tyutin embedding procedure. 

\noindent
%Keywords: symplectic, Hamilton-Jacobi, BRST symmetries\\
\vskip 0.5cm \noindent
\end{quotation}
%\end{center}
%-------------------------------------------------------------------------------------\\
%\noindent{\small Electronic address: sthong@ccs.sogang.ac.kr}
\par
\noindent \vskip 0.5cm \noindent
\end{titlepage}

\newpage

%%%%%%%%%%%%%%%%%%%%%%%%%%%%%%%%%%%%%%%%%%%%%%%%%%%%%%%%%%%%%%%%%%%%%%%%%%%%%
\section{Introduction}
%%%%%%%%%%%%%%%%%%%%%%%%%%%%%%%%%%%%%%%%%%%%%%%%%%%%%%%%%%%%%%%%%%%%%%%%%%%%%

Since the (2+1) dimensional O(3) nonlinear sigma model (NLSM) was
first discussed~\cite{polyakov75,phyrep}, there have
been many attempts to improve this soliton model associated
with the homotopy group $\pi_{2}(S^{2})=Z$.  Moreover,
the O(3) NLSM with the Hopf term~\cite{wilczek82}, as well as the $CP^{1}$ model associated with
the O(3) NLSM via the Hopf map projection from $S^{3}$ to $S^{2}$, were canonically
quantized~\cite{bowick86,pan88,kovner89} in the standard Dirac formalism~\cite{dirac64}.
Furthermore, the geometrical constraints involved in
the soliton models such as the Skyrmion models~\cite{sk2,wong}, O(3) NLSM~\cite{hong99o3}
and $CP^{1}$ model~\cite{hong00cp} have recently been analyzed in the
framework of the Batalin-Fradkin-Tyutin (BFT) scheme~\cite{BFT,BFT1,kpr}, to obtain the
corresponding first-class Hamiltonians and BRST~\cite{brst} invariant effective
Lagrangians~\cite{bfv,fik}.

On the other hand, an alternative symplectic scheme~\cite{jackiw85} has been proposed for treating 
first-order Lagrangians.  This scheme has been applied to various models~\cite{wozneto,kimjkps2,RR,neto0109}
and has recently been extended to include BRST invariant embeddings~\cite{HKPR}.  Moreover, since
the Hamilton-Jacobi (HJ) scheme~\cite{car} has first been  applied to constrained
systems~\cite{guler89}, there have been further developments in this direction
to quantize singular systems~\cite{pimentel98,baleanu01,hong01ph}.  In particular, the HJ description of 
second-class constrained systems, which posed some problems in previous
work~\cite{gomis}, has been shown to be complete after addition of suitable integrability
conditions, corresponding to the Dirac consistency
conditions, requiring time independence of the constraints~\cite{hong01ph}.  Recently
the O(3) NLSM was also considered in an extended HJ scheme~\cite{baleanu01} 
based on the extension of the second-class fields to the first-class fields  previously
introduced in ref.~\cite{hong99o3}.
However, this extension of the HJ scheme continues to involve some subtleties
still to be resolved.

In this paper, we will revisit the O(3) NLSM and examine the relation among the Lagrangian, symplectic, 
HJ and BFT embedding schemes for this model.
In section 2 we will analyze the constraint structure of this model from the point of view of the 
Lagrangian and symplectic algorithm. In section 3, we will investigate the HJ quantization of the O(3)
NLSM taking account of the integrability condition.  In section 4 we then discuss the BFT quantization scheme for this model, and
conclude in section 5.

%%%%%%%%%%%%%%%%%%%%%%%%%%%%%%%%%%%%%%%%%%%%%%%%%%%%%%%%%%%%%%%%%%%%%%%%%%
\section{NLSM in Lagrangian and symplectic schemes}
\setcounter{equation}{0}
\renewcommand{\theequation}{\arabic{section}.\arabic{equation}}
%%%%%%%%%%%%%%%%%%%%%%%%%%%%%%%%%%%%%%%%%%%%%%%%%%%%%%%%%%%%%%%%%%%%%%%%%%%%

In this section, we consider the Lagrangian and symplectic algorithms to generate the constraint structure of the O(3) NLSM,  defined by the Lagrangian
\begin{equation}
{\cal L}= \frac{1}{2f}(\pa_{\mu}n^{a})(\pa^{\mu}n^{a})
+n^{0}(n^{a}n^{a}-1),
\label{L-quadratic}
\end{equation}
where $n^{a}$ $(a=1,2,3)$ is a multiplet of three real scalar fields which parameterize an internal space
$S^{2}$, and $n^{0}$ is a Lagrange multiplier field implementing the geometrical constraint $n^{a}n^{a}-1=0$. From the Lagrangian (\ref{L-quadratic}) the canonical momenta conjugate to the field $n^{0}$ and the real scalar 
fields $n^{a}$ are given by 
\beq 
\label{momenta} 
\pi^{0}=0,~~~\pi^{a}=\frac{1}{f}\partial_0{n}^{a}, 
\eeq 
to yield the canonical Hamiltonian
\begin{equation}
{\cal H}^{(0)}=\frac{f}{2}\pi^{a}\pi^{a}+\frac{1}{2f}(\na n^{a})^{2}-n^{0}(n^{a}n^{a}-1).
\label{canH}
\end{equation}

Following ref.~\cite{jackiw85}, we rewrite the second-order Lagrangian (\ref{L-quadratic}) in first-ordered form as
\begin{equation}
{\cal L}^{(0)} = \pi^{a}\partial_0{n}^a - {\cal H}^{(0)}, \label{L-firstorder}
\end{equation}
where ${\cal L}^{(0)}$ is to be treated as a function of the {\it Lagrangian} variables
($n^0,n^{a},\pi^{a}$), and ${\cal H}^{(0)}$ is the canonical Hamiltonian (\ref{canH}).  
The original second-order formulation is recovered on shell, upon using the equation of motion for $\pi^a$.
Note that since ${\cal H}^{(0)}$ depends on $\pi^{a}$ and $n^{a}$, but not on their time derivatives, 
it can be regarded as the (level zero) symplectic potential. The Lagrangian is of the form
\begin{equation}
{\cal L}^{(0)} = a_{\alpha}\partial_0{\xi}^{\alpha} - {\cal H}^{(0)},
\label{genL}
\end{equation}
where we have denoted the set of the symplectic variables
$(n^{a},\pi^{a},n^{0})$ collectively by $\xi^{\alpha}$ and the corresponding conjugate 
momenta by $a_{\alpha}$. The dynamics of the model is then governed by the symplectic two-form matrix:
\begin{equation}
F^{(0)}_{\alpha\beta} (x,y) = \frac{\partial a_\beta(y)}
             {\partial\xi^{\alpha}(x)} -
         \frac{\partial a_\alpha(x)}{\partial \xi^\beta(y)},
\end{equation}
via the equations of motion
%%%%%
\footnote{For notational convenience we will omit component indices for vectors 
and matrices except cases where confusion may arise.}
%%%%%
\begin{equation}\label{EL}
\int{\rm d}^2y~F^{(0)}(x,y)\partial_0{\xi}(y)=K^{(0)}(x),
\end{equation}
where
\begin{equation}
K^{(0)}_\alpha(x) =  \frac{\delta}{\delta \xi^\alpha(x)}
\int {\rm d}^2 y {\cal H}_{0}(y)\nonumber\\
= \left( 
\begin{array}{c}
-2n^0(x)n^{a}(x) -\frac{1}{f}\vec\nabla^{2} n^a(x)\\
f\pi^a(x)\\
-n^a(x)n^a(x) + 1\\
\end{array}
\right).
\end{equation}

In the O(3) NLSM the symplectic two-form matrix is given by
\begin{equation}
F^{(0)}(x,y) =
\left(
\begin{array}{ccc}
O &-I &\vec 0  \\
I &O  &\vec 0 \\
\vec 0^T &\vec 0^T  &0  \\
\end{array}
\right)
\delta^2(x-y),
\end{equation}
where $O$ and $I$ stand for $3\times 3$ null and identity matrices, respectively, and the superscript $T$ denotes ``transpose". 
The zero-level symplectic two-form
has a zero mode
%%%%%%
\footnote{The superscript $T$ on the vectors is implied}
%%%%%%
${\nu}^{(0)}_{y}(x)=(\vec 0,\vec 0,1)\delta^2 (x-y)$, which generates the constraint
\begin{equation}
\int d^2y~\nu_{y}^{(0)}(x)K^{(0)}(y) =-\Omega_{1}(x)=0
\end{equation}
with
\begin{equation}
\Omega_{1}=n^{a}n^{a} - 1.
\label{Omega1}
\end{equation}
From here on we may proceed in two ways.

\bigskip\noindent
%%%%%%%%%%%%%%%%%%%%%%%%%%
i) {\it Lagrangian algorithm}
%%%%%%%%%%%%%%%%%%%%%%%%%%
\bigskip

In the Lagrangian  algorithm, we add the time derivative of the constraint $\Omega_{1}=0$ 
to the Lagrange equations of motion (\ref{EL}). This leads to the following enlarged set of
equations (for a more detailed discussion see ref.~\cite{RR}),
\begin{equation}
\int d^2y~W^{(1)}(x,y)\partial_0\xi(y) = K^{(1)}(x)
\end{equation}
where $W^{(1)}(x,y)$ are now the elements
of a {\it rectangular} matrix 
\begin{equation}
W^{(1)}(x,y)=\left(
\begin{array}{c}
F^{(0)}(x,y)\\
M_{1}(x,y)
\end{array}\right)
\end{equation}
with
\begin{equation}\label{M1}
M_{1\alpha}(x,y)= 
\frac{\partial \Omega_1(x)}{\partial \xi^\alpha(y)}
= (2\vec n(x),\vec 0,0)\delta^2(x-y)
\end{equation}
and
\begin{equation}\label{K1}
K^{(1)}(x)=\left(
\begin{array}{c}
K^{(0)}(x)
\\
0
\end{array}\right).
\end{equation}
The matrix $W^{(1)}$ has two left zero modes, of which one just reproduces the
previous constraint, while the other
\beq
\nu^{(1)}_{y}(x)=(\vec 0, 2\vec n(x),0,-1)\delta^2(x-y)
\eeq
yields the new constraint
\begin{equation}\label{Omega2}
\Omega_2= n^a\pi^a=0.
\end{equation}
We proceed in this fashion, and add the time derivative of this constraint
to the previous equations of motion, which now read
\begin{equation}
\int d^2y~W^{(2)}(x,y)\partial_0\xi(y) = K^{(2)}(x)
\end{equation}
where $W^{(2)}(x,y)$ are now the elements
of a {\it rectangular} matrix 
\begin{equation}
W^{(2)}(x,y)=\left(
\begin{array}{c}
F^{(0)}(x,y)\\
M_{1}(x,y)\\
M_{2}(x,y)
\end{array}\right)\,\,,
\end{equation}
with
\begin{equation}\label{M2}
M_{2\alpha}(x,y)=\frac{\partial \Omega_2(x)}{\partial \xi^\alpha(y)}
= (\vec\pi(x), \vec n(x), 0)\delta^2(x-y)
\end{equation}
and
\begin{equation}\label{K2}
K^{(2)}(x)=\left(
\begin{array}{c}
K^{(0)}(x)\\
0\\
0
\end{array}\right).
\end{equation}
Now, $W^{(2)}$ is found to have three zero modes, of which 
two reproduce the previous constraints, while the zero mode
\begin{equation}
\nu_{y}^{(2)}(x)= (\vec n(x), -\vec\pi(x), 0,0,1)\delta^2(x-y)
\end{equation}
leads to a further constraint:
\begin{equation}\label{Omega3}
\Omega_3 = 2n^0 n^an^a + f\pi^a\pi^a 
+\frac{1}{f}n^a\vec\nabla^{2} n^a=0.
\end{equation}
Extending the matrix $W^{(2)}$ to  $W^{(3)}$ in the manner described
above, we find that there are {\it no} new zero modes, and the iterative algorithm
terminates at this point. We have thus four constraints, of which
two constraints $\Omega_i=0$ $(i=1,2)$ represent constraints among the variables $n^a$ and $\pi^a$ alone.

As one readily checks, the constraints are found to completely agree with the constraints 
generated by the conventional Dirac algorithm, departing either from the original second-order 
Lagrangian (\ref{L-quadratic}) or from the first-order Lagrangian (\ref{L-firstorder}).

\bigskip\noindent
%%%%%%%%%%%%%%%%%%%%%%%%%%
ii) {\it Symplectic algorithm}
%%%%%%%%%%%%%%%%%%%%%%%%%%
\bigskip

In the symplectic algorithm of ref. \cite{wozneto} we add the constraint $\Omega_{1}$ to the 
canonical sector of the Lagrangian (\ref{L-firstorder}) in the form
%%%%%
\footnote{The minus sign is included in order to align our results with those
of the Lagrangian algorithm.}
%%%%% 
$-\Omega_1\partial_{0}\rho$, thereby enlarging the symplectic phase 
space via the addition of a dynamical field $\rho$, which
enforces the stability of the constraint $\Omega_1$ in time. The once iterated first-level Lagrangian is then
given as follows
\begin{equation}
{\cal L}^{(1)} = \pi^{a}\partial_0{n}^{a}-\Omega_{1}\partial_0{\rho}-{\cal H}^{(0)}.
\label{lagsecond}
\end{equation}
In the spirit of ref. \cite{wozneto} we absorb the Lagrange multiplier
field $n^0$ in ${\cal H}^{(0)}$ into the new dynamical variable
$\rho$ by making the substitution $\partial_0\rho - n^0 \rightarrow \partial_0\rho$.
This amounts to the replacement
\begin{equation}
{\cal L}^{(1)}\rightarrow {\cal L}^{(1)} = \pi^{a}\partial_0{n}^{a}
-\Omega_{1}\partial_0{\rho}-{\cal H}^{(1)}.
\label{lagsecond2}
\end{equation}
where ${\cal H}^{(0)}$ in Eq. (\ref{lagsecond}) has been replaced by
\begin{equation}
{\cal H}^{(1)}=\frac{f}{2}\pi^{a}\pi^{a}+\frac{1}{2f}(\na n^{a})^{2}.
\label{hamilsecond}
\end{equation}
Note that at this stage the field $n^0$ has completely disappeared, and any information concerning 
it has been lost. In fact, the model that we are discussing at this point is the model considered 
in ref. \cite{baleanu01}, and hence is no longer the O(3) NLSM in that sense.

The new equations of motion now read
\beq
\int d^2y~F^{(1)}(x,y) \partial_0\xi^{(1)}(y) 
= K^{(1)}(x)=
\left(
\begin{array}{c}
-\frac{1}{f}\vec\nabla^2 n^a (x)\\
f\pi^a (x)\\
0
\end{array}
\right)
\eeq
where
\begin{equation}
\xi^{(1)}(x)=(\vec n(x),\vec\pi(x),\rho(x))
\end{equation}
and the first-iterated symplectic two-form is now given by
\begin{equation}
F^{(1)}(x,y) =
\left( \
\begin{array}{ccc}
         O &-I   &-2\vec n(x) \\
         I &O    &\vec 0 \\
       2\vec n(x)^T&\vec 0^T& 0
\end{array}
\right)
\delta^2 (x-y).
\end{equation}
Note that the symplectic two-form matrix still has a
zero mode $\nu^{(1)}_{y}(x)=(\vec 0,2\vec n(x),1)\delta^2 (x-y)$, which generates the constraint $\Omega_{2}$ in the
context of the symplectic formalism~\cite{jackiw85} as follows
\begin{equation}
\int d^2y~\nu^{(1)}_{y}(x)K^{(1)}(y) = 2f\Omega_{2}(x)=0\,,
\end{equation}

Following ref.~\cite{wozneto} we further enlarge the symplectic phase space to include the constraint $\Omega_2$ via an
additional dynamical field $\sigma$, as follows,
\begin{equation}
{\cal L}^{(2)} = \pi^{a}\partial_0{n}^{a}-\Omega_{1}\partial_0{\rho}-\Omega_{2}\partial_0 {\sigma}-{\cal H}^{(2)}, 
\label{lagthird}
\end{equation}
with ${\cal H}^{(2)}={\cal H}^{(1)}$. The
symplectic variables $\xi^{(2)}$ are now given by
\begin{equation}
\xi^{(2)}(x)=(n^{a}(x),\pi^{a}(x),\rho(x),\sigma(x)),
\end{equation}
and the second-iterated symplectic two-form matrix is given by
\begin{equation}
F^{(2)}(x,y) = \left( \
\begin{array}{cccc}
         O &-I   &-2\vec n(x) &-\vec\pi(x) \\
         I &O    &\vec 0 &-\vec n(x) \\
       2\vec n(x)^T&\vec 0^T& 0 &0 \\
       \vec \pi(x)^T& \vec n(x)^T& 0 & 0
\end{array}
\right) \delta^2 (x-y).
\end{equation}
We easily show that this symplectic matrix is invertible, i.e no zero modes exist. Hence the
algorithm ends at this point.

Although the constraints $\Omega_{i}=0$ $(i=1,2)$ are identical with those generated by 
the Lagrangian algorithm, a complete analysis of the NLSM should have treated
the Lagrange multiplier field $n^0$ as an additional degree of freedom,
by not performing the shift in variable $\partial_0\rho - n^0 \rightarrow\partial_0\rho$. 
As shown in ref. \cite{RR} the symplectic algorithm described
above does not generate in this case the complete set of constraints as given
by the Lagrangian algorithm, which rests on a solid foundation. The reason for 
this failure has been examined in ref. \cite{RR}
for the corresponding quantum mechanical model, as well as in general terms.

%%%%%%%%%%%%%%%%%%%%%%%%%%%%%%%%%%%%%%%%%%%%%%%%%%
\section{NLSM in Hamilton-Jacobi scheme}
\setcounter{equation}{0}
\renewcommand{\theequation}{\arabic{section}.\arabic{equation}}
%%%%%%%%%%%%%%%%%%%%%%%%%%%%%%%%%%%%%%%%%%%%%%%%%%%%%

In this section we revisit the O(3) NLSM in the HJ scheme, 
for which the generalized
HJ partial differential equations are given by
%%%%%%%
\footnote{This scheme was treated incompletely in ref.~\cite{baleanu01}.
With the identification $\pi^a(x) = \frac{\delta S}{\delta n^a(x)}$ and $p^0 =\frac{\delta S}{\delta t}$, 
these equations are the (incomplete set of)
Hamilton-Jacobi partial differential equations for the Hamilton principal function $S$.}
%%%%%%%
\bea
\label{hj}
{\cal H}^{\prime}_0 &=&p^{0} +{\cal H}_{0}=0,\nn\\
{\cal H}^{\prime}_1 &=&\pi^{0} + {\cal H}_{1}=0,
\eea
where ${\cal H}_{0}$ is a canonical Hamiltonian density ${\cal H}^{(0)}(n^0,n^a,\pi^a)$ 
in Eq. (\ref{canH}) and ${\cal H}_{1}^{\pr}$
is the (only) primary constraint in the Dirac terminology~\cite{dirac64} 
(${\cal H}_{1}$ is actually zero for the case of the O(3) NLSM, as seen from 
Eq. (\ref{momenta})).
 
Following the generalized HJ scheme~\cite{HKPR,hong01ph}, one obtains from Eq. (\ref{hj}),
\bea
dq^{\un{a}}&=&\frac{\pa {\cal H}^{\prime}_{\un{\alpha}}}
{\pa p^{\un{a}}}dt_{\un{\alpha}},\nn\\
dp^{\un{a}}&=&-\frac{\pa {\cal H}^{\prime}_{\un{\alpha}}}
{\pa q^{\un{a}}}dt_{\un{\alpha}},
\label{hjem0}
\eea
where $q^{\un{a}}=(t,n^{0},n^{a})$, $p^{\un{a}}=(p^{0},\pi^{0},\pi^{a})$
and $t_{\un{\alpha}}=(t,n^{0})$ with $\un{\alpha} = (0,1)$. Explicitely
\begin{eqnarray}
&&\pa_{0} n^{0}=\pa_{0} n^{0},~~~~~~~~~\pa_{0} n^{a}=f\pi^{a}, \nonumber \\
&&\pa_{0}\pi^{0}=n^{a}n^{a}-1,
   ~~~\pa_{0}\pi^{a}=\frac{1}{f}\na^{2}n^{a}+2n^{0}n^{a}.
\label{da0dpi0}
\end{eqnarray}
Since the equation for $n^{0}$ is trivial, one cannot obtain any
information about the variable $n^{0}$ at this level, and the set of equations
is not integrable at this stage.

In order to fix this deficit of information, we supplement Eq. (\ref{hj}) 
with the generalized integrability conditions~\cite{HKPR,hong01ph},
\beq
\partial_{0}{\cal H}^{\prime}_{\bar{\alpha}}
=\{{\cal H}^{\prime}_{\bar{\alpha}}, {\cal H}^{\prime}_{0}\}
+\{{\cal H}^{\prime}_{\bar{\alpha}}, {\cal H}^{\prime}_{\beta}\}
\partial_{0}q^{\beta}=0,
\label{integ2}
\eeq
where, unlike in the usual case, the Poisson bracket is defined in terms of the extended 
index $\un{a}$ corresponding to $q^{\un{a}}=(t,n^{0},n^{a})$ as follows
\begin{equation}
\label{partial}
\{A,B\}
=\frac{\pa A}{\pa q^{\un{a}}}\frac{\pa B}
{\pa p^{\un{a}}}
-\frac{\pa B}{\pa q^{\un{a}}}\frac{\pa A}
{\pa p^{\un{a}}}\,.
\end{equation}
The index $\beta$ in (\ref{integ2}) labels the primary constraints, whose 
number is only one for the case in question ($\partial_0 q^\beta \to \partial_0 n^0$).
The index $\bar\alpha$, on the other hand, labels the primary constraints as well 
as the secondary constraints which emerge {\it iteratively} from the 
integrability condition (\ref{integ2}). Thus for the case in question we have
three secondary constraints,  ${\cal H}_{2}^{\pr}=0$,
${\cal H}_{3}^{\pr}=0$ and ${\cal H}_{4}^{\pr}=0$ corresponding to  
Eqs. (\ref{Omega1}), (\ref{Omega2}) and (\ref{Omega3}), respectively.
They emerge iteratively from Eq. (\ref{integ2}) as follows:
\bea
\partial_{0}{\cal H}^{\prime}_0 &=&- {\cal H}^{\prime}_2 \partial_{0}n^{0}\,,\nn\\
\partial_{0}{\cal H}^{\prime}_1 &=& {\cal H}^{\prime}_2\,,\nn\\
\partial_{0}{\cal H}^{\prime}_2 &=&2f{\cal H}^{\prime}_3\,,\nn\\
\partial_{0}{\cal H}^{\prime}_3 &=&{\cal H}^{\prime}_4\,,\nn
\eea
with
\bea
{\cal H}^{\prime}_2&=&n^{a}n^{a}-1\,,
\nn\\
{\cal H}^{\prime}_{3}&=&n^{a}\pi^{a}\,.
\label{hpr}\\
{\cal H}_4^\prime &=&2n^0 n^an^a 
+ f\pi^a\pi^a +\frac{1}{f}n^a\vec\nabla^2n^a.
\nn
\eea
From $\partial_0 {\cal H}^\prime_4 = 0$ we have,
using ${\cal H}^\prime_2 = {\cal H}^\prime_3 = 0$ as well as (\ref{da0dpi0}),
\beq
\pa_{0} n^{0}=
-\pi^{a}\na^{2}n^{a}+\na n^{a}\cdot\na\pi^{a}
\label{dn0dt}
\eeq
which replaces the identity $\pa_{0}n^0=\pa_{0}n^0$ in Eq. (\ref{da0dpi0}), 
and thus completes the set of Hamilton-Jacobi equations to an integrable system.

Finally note that Eqs. (\ref{hjem0}) and (\ref{hpr}) imply
for the Hamilton principal function
\begin{eqnarray}
\label{hjem1}
dS&=&\int{\rm d}^{2}x~\left(-{\cal H}_{\un{\alpha}}
+\pi^{a}\frac{\partial{\cal H}^{\prime}_{\un{\alpha}}}
{\partial\pi^{a}}\right)dt_{\un{\alpha}},\nn\\
&=&dt\int{\rm d}^{2}x~ \left(-{\cal H}_{0}
+\pi^{0}\frac{\partial n^{0}}{\partial t}
+\pi^{a}\frac{\partial n^{a}}{\partial t}\right).
\end{eqnarray}
Since we have now from Eqs. (\ref{da0dpi0}) and (\ref{dn0dt}) a complete set of 
equations of motion for $n^{a}$ and $n^{0}$, Eq. (\ref{hjem1}) can
be integrated in time to yield the standard action
\beq
S=\int{\rm d}^{3}x~{\cal L}^{(0)}
\label{action0fin}
\eeq
where ${\cal L}^{(0)}$ is the first-order Lagrangian (\ref{L-firstorder}). Note that 
it was only after taking account of the secondary constraints generated
iteratively by the integrability
condition (\ref{integ2}), that one could construct the action
(\ref{action0fin}) for the second-class system in question.

%%%%%%%%%%%%%%%%%%%%%%%%%%%%%%%%%%%%%%%%%%%%%%%%%%
\section{BFT Hamiltonian embedding}
\setcounter{equation}{0}
\renewcommand{\theequation}{\arabic{section}.\arabic{equation}}
%%%%%%%%%%%%%%%%%%%%%%%%%%%%%%%%%%%%%%%%%%%%%%%%%%%%%

In this section  we reconsider the BFT quantization for the O(3) NLSM
with the Lagrangian (\ref{L-quadratic}) and the canonical Hamiltonian (\ref{canH}). 
\footnote{In  previous work~\cite{hong99o3} we have carried out the BFT embedding
for the O(3) NLSM without explicitly including the geometrical constraint $n^{a}n^{a}-1=0$
with the Lagrangian multiplier field $n^{0}$ in the starting Lagrangian.} The primary constraint reads
\begin{equation}
\Omega_0=\pi^{0} \approx 0,
\label{const0}
\end{equation}
and correspondingly we have for the total Hamiltonian,
\beq\label{H-total}
{\cal H}_{T} = {\cal H}_{0} + v\Omega_0,
\eeq
where ${\cal H}_{0}$ is the canonical Hamiltonian ${\cal H}^{(0)}$ in Eq. (\ref{canH}).  The usual Dirac 
algorithm is readily shown to lead recursively to the constraints 
$\Omega_{i}$ $(i=0,1,2,3)$ as follows
%%%%%%%
\footnote{Note that these constraints correspond to
${\cal H}_{i}^{\prime}=0$ $(i=1,2,3,4)$ in the HJ scheme, respectively.}
%%%%%%% 
\begin{eqnarray}
\{\Omega_0(x),{\cal H}_0(y)\}&=&\Omega_1(x)\delta^{2}(x-y),\nonumber\\
\{\Omega_1(x),{\cal H}_0(y)\}&=&2f\Omega_2(x)\delta^{2}(x-y),\nonumber\\
\{\Omega_2(x),{\cal H}_0(y)\}&=&\Omega_3(x)\delta^{2}(x-y).
\label{recursive}
\end{eqnarray}
The requirement of time-independence of $\Omega_3$ then finally fixes the Lagrange 
multiplier field in Eq. (\ref{H-total}) to $v= 0$. Notice that for the case of the 
motion of a particle on a sphere, the constraint $\Omega_3=0$ just fixes the ``string tension"
of the central force governing its motion.
 
The complete system of constraints is thus second-class. 
The subset $\Omega_1=0$ and $\Omega_2=0$  only links the fields $n^a$ and $\pi^a$, while the
remaining constraints $\Omega_{0}=0$ and $\Omega_{3}=0$ link these variables to the rest 
of the variables. We may thus ``gauge" the $(n^a,\pi^a)$ sector a la BFT, while 
leaving a strong implementation of the
constraints $\Omega_0=0$ and $\Omega_3=0$
to the end. This is possible since the corresponding Dirac brackets,
constructed from the inverse of the matrix
\beq
\{\Omega_{1}(x),\Omega_{2}(y)\}= \left(
\begin{array}{cc}
0&2n^{a}(x)n^{a}(x)\\
-2n^{a}(x)n^{a}(x)&0
\end{array}
\right)\delta^{2}(x-y),
\eeq
reduce to
ordinary Poisson brackets for functionals of $n^a$ and $\pi^a$ alone.

Following the BFT scheme~\cite{BFT,BFT1,kpr}, we systematically
convert the second-class constraints $\Omega_i=0$ $(i=1,2)$ into first-class ones
by introducing two canonically conjugate auxiliary fields 
$(\theta, \pi_{\theta})$  with Poisson brackets
\begin{equation}
\{\theta(x), \pi_{\theta}(y)\}=\delta^{2}(x-y).
\label{phii}
\end{equation}
The strongly involutive first-class constraints $\tilde{\Omega}_{i}$ are then constructed as a
power series of the auxiliary fields~\cite{hong99o3},
\begin{eqnarray}
\tilde{\Omega}_{1}&=&\Omega_{1}+2\theta,  \nonumber \\
\tilde{\Omega}_{2}&=&\Omega_{2}-n^{a}n^{a}\pi_{\theta},
\label{1stconst}
\end{eqnarray}
which satisfy the closed algebra $\{\tilde{\Omega}_{i},\tilde{\Omega}_{j}\}=0$.

We next construct the first-class fields $\tilde{{\cal F}}
=(\tilde{n}^{a},\tilde{\pi}^{a})$, corresponding to the original fields
defined by ${\cal F}=(n^{a},\pi^{a})$ in the extended phase space. They are
obtained as a power series in the auxiliary fields $(\theta,\pi_{\theta})$
by demanding that they be in strong involution with the first-class constraints
(\ref{1stconst}), that is
$\{\tilde{\Omega}_{i}, \tilde{{\cal F}}\}=0$.  After some tedious algebra, we 
obtain for 
the first-class physical fields
\begin{eqnarray}
\tilde{n}^{a}&=&n^{a}\left(\frac{n^{a}n^{a}+2\theta}{n^{a}n^{a}}\right)^{1/2},\nonumber \\
\tilde{\pi}^{a}&=&\left(\pi^{a}-n^{a}\pi_{\theta}\right)\left(\frac{n^{a}n^{a}}
{n^{a}n^{a}+2\theta}\right)^{1/2}.
\label{pitilde}
\end{eqnarray}
They are found to satisfy the Poisson algebra
\begin{eqnarray}
\{\tilde{n}^{a}(x),\tilde{\pi}^{b}(y)\}&=&(\delta^{ab}-\frac{\tilde{n}^{a}
\tilde{n}^{b}}{\tilde{n}^{c}\tilde{n}^{c}})\delta^{2}(x-y),  \nonumber \\
\{\tilde{\pi}^{a}(x),\tilde{\pi}^{b}(y)\}&=&\frac{1}{\tilde{n}^{c} \tilde{n}%
^{c}}(\tilde{n}^{b}\tilde{\pi}^{a} -\tilde{n}^{a}\tilde{\pi}^{b})\delta^{2}(x-y).  \label{commst}
\end{eqnarray}
To our knowledge, this is the first time that  the first-class fields in the 
O(3) NLSM have been given  in this compact form.
%%%%%%%%%%%%%
\footnote{Note that the first-class physical fields were obtained in
terms of the infinite power series of the auxiliary fields $(\theta,\pi_{\theta})$
in the previous work~\cite{hong99o3}.}
%%%%%%%%%%%%%
Note that in terms of the first-class fields (\ref{pitilde}) the constraints
(\ref{1stconst}) take the  natural form
\be\label{firstclassconstr}
\tilde\Omega_1 = \tilde n^a\tilde n^a - 1\,,\quad
\tilde\Omega_2 = \tilde n^a\tilde\pi^a\label{omegatilde}\,.
\ee
Eq. (\ref{firstclassconstr}) illustrates that  any functional ${\cal K}(\tilde{{\cal F}})$ of the first-class fields $\tilde{{\cal F}}$
is also first-class. We correspondingly construct the first-class Hamiltonian in terms of the above first-class
physical variables by making the replacements $n^a\to\tilde n^a,
\pi^a\to\tilde\pi^a$ in the canonical Hamiltonian 
${\cal H}_{0}$,
\begin{equation}
\tilde{\cal H}_{0}=\frac{f}{2}\tilde{\pi}^{a}\tilde{\pi}^{a} +
\frac{1}{2f}(\na\tilde{n}^{a})^{2}
-{n}^{0}(\tilde{n}^{a}\tilde{n}^{a}-1).
\label{htilde}
\end{equation}

Note that the first-class Hamiltonian (\ref{htilde}) 
is manifestly strongly involutive with the
first-class constraints, $\{\tilde{\Omega}_{i},\tilde{\cal H}_{0}\}=0$.
This need not be so. Indeed, 
this Hamiltonian is not unique, as we may always add to
it terms proportional to the first class constraints without altering the dynamics
of the first-class fields. Thus we could ask the gauged 
total Hamiltonian to generate the first-class constraints in an
analogous way to Eq. (\ref{recursive})
\begin{eqnarray}\label{recursive2}
\{\tilde\Omega_0(x),\tilde {\cal H}(y)\}&=&\tilde\Omega_1(x)\delta^{2}(x-y),\nonumber\\
\{\tilde\Omega_1(x),\tilde {\cal H}(y)\}&=&2f\tilde\Omega_2(x)\delta^{2}(x-y),\nonumber\\
\{\tilde\Omega_2(x),\tilde {\cal H}(y)\}&=&\Omega_3(x)\delta^{2}(x-y)\,.
\end{eqnarray}
This may be achieved via the replacement of (\ref{htilde}) by
\beq\label{Htilde-prime}
\tilde{\cal H}= \tilde{\cal H}_0 + f\pi_\theta \tilde{\Omega}_2
-\frac{1}{2}\ln(\tilde{n^a}\tilde{n^a}-2\theta)\Omega_{3}.
\eeq
Since the second class constraint $\Omega_3=0$ should eventually
be implemented strongly in our scheme, we shall set it equal to zero.
As for $\tilde{\Omega}_2$, it is first class, so that it will not
participate in the dynamics of gauge invariant functionals
in the $(n^0,\pi^0,n^a,\pi^a,\theta,\pi_\theta)$ space. This allows us to set the last two 
terms in Eq. (\ref{Htilde-prime}) equal to zero.  Rewriting the Hamiltonian (\ref{htilde}) 
in terms of the original fields in the $(n^a,\pi^a)$ sector, and 
the auxiliary ones in the $(\theta,\pi_{\theta})$ sector, we have  
\begin{eqnarray}
\tilde{\cal H}_{0}&=&\frac{f}{2}(\pi^{a}-n^{a}\pi_{\theta})
(\pi^{a}-n^{a}\pi_{\theta})\frac{n^{c}n^{c}}{n^{c}n^{c}+2\theta}
+\frac{1}{2f}(\na n^{a})^{2}\frac{n^{c}n^{c}+2 \theta}{n^{c}n^{c}}
\nonumber \\
& &-n^{0}(n^{a}n^{a}-1+2\theta),
\label{hct}
\end{eqnarray}
where we have used the conformal map condition, $n^{a}\na n^{a}=0$, which
states that the radial vector is perpendicular to the tangent on the $S^{2}$
sphere in the extended phase space of the O(3) NLSM~\cite{hong99o3}.
 
In the framework of the BFV formalism~\cite{bfv,fik}, 
we now construct the nilpotent BRST charge $Q$, the
fermionic gauge fixing function $\Psi$ and the BRST invariant minimal
Hamiltonian $H_{m}$  by  introducing two canonical
sets of ghost and anti-ghost fields, together with auxiliary fields
$({\cal C}^{i},\bar{{\cal P}}_{i})$, $({\cal P}^{i}, \bar{{\cal C}}_{i})$ and
$(N^{i},B_{i})$, $(i=1,2)$, 
\begin{eqnarray}
Q&=&\int {\rm d}^{2}x~({\cal C}^{i}\tilde{\Omega}_{i}+{\cal P}^{i}B_{i}),
\nonumber \\
\Psi&=&\int {\rm d}^{2}x~(\bar{{\cal C}}_{i}\chi^{i}+\bar{{\cal P}}%
_{i}N^{i}),  \nonumber \\
H_{m}&=&\int {\rm d}^{2}x~\left(\tilde{\cal H}_{0}-2f{\cal C}^{1}\bar{{\cal P}}_{2}\right),
\label{hmham}
\end{eqnarray}
with the properties $Q^{2}=\{Q,Q\}=0$ and $\{\{\Psi,Q\},Q\}=0$.
The nilpotent charge $Q$ is the generator of the following infinitesimal
transformations,
\beq
\begin{array}{lll}
\delta_{Q}n^{0}=0, &~~\delta_{Q}n^{a}=-{\cal C}^{2}n^{a},
&~~\delta_{Q}\theta={\cal C}^{2}n^{a}n^{a},\\
\delta_{Q}\pi^{0}=0,
&~~\delta_{Q}\pi^{a}=2{\cal C}^{1}n^{a}+{\cal C}^{2}(\pi^{a}-2n^{a}\pi_{\theta}),
&~~\delta_{Q}\pi_{\theta}=2{\cal C}^{1},\\
\delta_{Q}\bar{{\cal C}}_{i}=B_{i}, &~~\delta_{Q}{\cal C}^{i}=0, &~~\delta_{Q}B_{i}=0,\\
\delta_{Q}{\cal P}^{i}=0, &~~\delta_{Q}\bar{{\cal P}}_{i}=\tilde{\Omega}_{i},
&~~\delta_{Q}N^{i}=-{\cal P}^{i},\\
\end{array}
\label{brstgaugetrfm}
\eeq
which in turn imply $\{Q,H_{m}\}=0$, that is,  $H_{m}$ in Eq. (\ref{hmham})
is  the BRST invariant minimal Hamiltonian.
  
After some algebra, we arrive at the effective quantum Lagrangian of the manifestly covariant form
\begin{equation}
{\cal L}_{eff}= {\cal L}^{(0)} + {\cal L}^{WZ} + {\cal L}^{ghost}
\label{lagfinal}
\end{equation}
where ${\cal L}^{(0)}$ is given by Eq. (\ref{L-quadratic}) and
\begin{eqnarray}
{\cal L}^{WZ}&=&
\frac{1}{fn^{c}n^{c}}(\partial_{\mu}n^{a})(\partial^{\mu}n^{a}){\theta}-\frac{1}{2f(n^{c}n^{c})^{2}}\partial_{\mu}\theta\partial^{\mu}\theta
+2n^{0}\theta,\nonumber\\
{\cal L}^{ghost}&=&-\frac{1}{2f}(n^{a}n^{a})^{2}(B+2\bar{{\cal C%
}}{\cal C})^{2} -\frac{1}{n^{c}n^{c}}\partial_{\mu}\theta\partial^{\mu}B
+\partial_{\mu}\bar{{\cal C}}\partial^{\mu}{\cal C}.
\label{lagwz}
\end{eqnarray}
This Lagrangian is invariant under the BRST transformation
\beq
\begin{array}{lll}
\delta_{\epsilon}n^{0}=0, &~~\delta_{\epsilon}n^{a}=\epsilon n^{a}{\cal C}, &~~\delta_{\epsilon}\theta=-\epsilon
n^{a}n^{a}{\cal C},\\
\delta_{\epsilon}\bar{{\cal C}}=-\epsilon B, &~~\delta_{\epsilon}{\cal C}=0, &~~\delta_{\epsilon}B=0,\\
\end{array}
\eeq
where $\epsilon$ is an infinitesimal Grassmann valued parameter.  Note that 
the Wess-Zumino Lagrangian in (\ref{lagwz}) involves $n^{0}$, which originates from the geometrical constraint in the starting Lagrangian.

%%%%%%%%%%%%%%%%%%%%%%%%%%%%%%%%%%%%%%%%%%%%%%%%%%%%%%%%%%%%%%%%%%%%%%%%%%
\section{Conclusion}
%%%%%%%%%%%%%%%%%%%%%%%%%%%%%%%%%%%%%%%%%%%%%%%%%%%%%%%%%%%%%%%%%%%%%%%%%

We have investigated the constraint structure of the O(3) NLSM, in the Lagrangian, 
symplectic, Hamilton-Jacobi (HJ) and Batalin-Fradkin-Tyutin (BFT) quantization schemes. In particular we showed that the symplectic algorithm led to an incomplete constraint structure for this model.
In fact, the missing constraint (\ref{Omega3}) was shown to play an essential role as a part of the integrability conditions in the Hamilton-Jacobi
formulation, necessary to recover the action for this second class system. 

We further considered the gauge-embedding of the NLSM following the
BFT quantization scheme for constructing the first-class fields
and first-class Hamiltonian including the Wess-Zumino terms.
A similar study, based on the ``first-class field approach" was done in ref. ~\cite{baleanu01} 
in the framework of the Hamilton-Jacobi scheme.
Even though the
authors of  ~\cite{baleanu01} exploited
the first-class Hamiltonian in a form similar to $\tilde{\cal H}_{0}$ in Eq. (\ref{htilde}), their final
Lagrangian is not covariant, in contrast to the covariant Lagrangian (\ref{lagfinal}) involving the 
Wess-Zumino and
ghost fields. Moreover the ``integrability conditions" were not taken into account.
The attempt of achieving such an embedding for the NLSM in the symplectic approach based on a biased 
``educated guess" for the general
Lorentz covariant 
form of the first-class gauged Lagrangian, as was done in refs. \cite{HKPR}
for the Proca and self-dual models, 
seems very problematic as seen from Eq. (\ref{lagwz}), which involves
inverse powers of $(n^{a}n^{a})$. We thus believe that for
the case of the O(3) NLSM, the BFT embedding procedure is the most tractable one as compared to the Lagrangian, symplectic and HJ schemes.

\vskip 1.0cm
STH would like to thank the warm hospitality of the Institut f\"ur Theoretische Physik at the Universit\"at 
Heidelberg where a part of this work has been done, and G\" uler and Baleanu for helpful discussions.  
The work of STH and YJP was supported by the Korea Research Foundation, Grant No. KRF-2001-DP0083.

%\begin{references}

\end{document}